\documentclass[twocolumn,showpacs,preprintnumbers,amsmath,amssymb,pre]{revtex4-1}
\usepackage{color}    
\usepackage{graphicx}
\usepackage{dcolumn}
\usepackage{bm}
\usepackage{subfigure}
\usepackage{amssymb}
\usepackage{multirow}
\usepackage{amsmath}
\usepackage{csquotes}
\graphicspath{{plots/}}

\newcommand\xyplane{$k$\nobreakdash--$\omega$~plane}

\renewcommand{\vec}[1]{\mathbf{#1}}

\begin{document}

\title{Theoretical foundations of quantum hydrodynamics for plasmas}

\author{ Zh.~A.  Moldabekov$^{1, 2, 3}$, M. Bonitz$^{1}$, and T. S.  Ramazanov$^{2, 3}$}

\affiliation{
 $^1$Institut f\"ur Theoretische Physik und Astrophysik, Christian-Albrechts-Universit\"at zu Kiel,
 Leibnizstra{\ss}e 15, 24098 Kiel, Germany}
 \affiliation{
 $^2$Institute for Experimental and Theoretical Physics, Al-Farabi Kazakh National University, 71 Al-Farabi str.,  
  050040 Almaty, Kazakhstan
  }
\affiliation{$^3$Institute of Applied Sciences and IT, 40-48 Shashkin Str., 050038 Almaty, Kazakhstan}

\begin{abstract}
Beginning from the semiclassical Hamiltonian, the Fermi pressure and Bohm potential for the quantum hydrodynamics application (QHD) at finite temperature are consistently derived  in  the  framework  of  the  local  density  approximation with the first order density gradient correction.
Previously known results are revised and improved with a clear description of the underlying approximations.
A fully non-local Bohm potential, which goes beyond of all previous results and is linked to the electron polarization function
 in the random phase approximation, for the QHD model is presented. The dynamic QHD exchange correlation potential is introduced in the framework of local field corrections, and considered for the case of the relaxation time approximation.
 Finally, the range of applicability of the QHD is discussed.
\end{abstract}

\pacs{xxx}

\maketitle
\section{Introduction}

The investigation of dynamical properties of systems containing partially or fully degenerate electrons has gained growing interest due
to their relevance for such fields like  dense plasmas \cite{kremp_springer_2005, Daligault}, warm dense matter \cite{Fletcher,Dornheim}, streaming and wake effects \cite{Moldabekov_PRE, moldabekov_cpp_2016} and plasmonics \cite{PRL2012, Ciraci, 52, 53}. 
Interiors of giant planets, white and brown dwarfs, stellar cores, and the outer envelope of neutron stars  are examples of matter in the state of a partially degenerate  dense plasma (warm dense matter) \cite{Fortney, Potekhin}.  Experimental studies of dense plasmas include the free electron laser excited plasmas \cite{Zastrau}, and inertial confinement fusion experiments \cite{Hurricane, Cuneo, Gomez}. On the other hand, plasmonic materials, containing fully degenerate electrons, 
  have recently received renewed attention as the result of the advances in nanofabrication techniques \cite{Chen, Chen2, 52, 53}.
  All the above mentioned different systems are governed by the behavior of Coulomb interacting quantum electrons.

  The theoretical description of quantum plasmas must take into account quantum degeneracy effects such as non-locality, spin statistics, and correlations (non-ideality) appropriately on the relevant scales. 
  A continuum description of the dynamics of the quantum electrons in the spirit of a density functional theory (DFT) is a promising approach to the problem. 
  The possibility of such a description follows from the Hohenberg-Kohn theorem of DFT \cite{Hohenberg, Kohn} and the 
  Runge-Gross theorem of time-dependent (or current) density functional theory (TDDFT) \cite{Runge}. The dynamics of the electrons can be described in terms of the average electron density $n({\vec r}, t)$ and 
  the average electron current density ${\vec j}({\vec r}, t)$ by the continuity and momentum equations \cite{Runge, Giuliani}:
  \begin{align}\label{TDFT_1}
\frac{\partial}{\partial t} n\left(\vec r, t\right)+\vec{\nabla} \left[\vec j \left(\vec r, t\right)\right] &=0,
\\
m \frac{\partial}{\partial t} \vec j\left(\vec r, t\right) - n \left(\vec r, t\right) e\vec E (\vec r, t) &= -\vec{\nabla} \cdot \vec P (\vec r, t),
\label{TDFT_2}
 \end{align}
  where $\vec E$ is the electric-field strength, and $\vec P$ stands for a tensor that contains all many-body and quantum effects, including correlations and dissipation.
  The exact form of $\vec P$ is not known, and different approaches and approximations for $\vec P$ have been proposed with different levels of sophistication, e.g. \cite{Tokatly, Gao}. 
  
An alternative concept are first-principle approaches based on wave function methods, e.g. \cite{hochstuhl_epjst_2014}, quantum-statistical theory \cite{kremp_springer_2005}, quantum kinetic theory \cite{Bonitz_book, bonitz_cpp_1999} or non-equilibrium Green functions \cite{balzer_springer_2013, schluenzen_cpp_2016}.
%as well as TDDFT 
However, as TDDFT, these methods require substantial theoretical and computational effort and are particularly important to capture electronic correlations.
Therefore, in cases where correlations and their dynamics are of minor importance, simpler approaches are being used. This particularly applies to quantum hydrodynamics (QHD) that became popular as a simplified approach for quantum plasmas \cite{Manfredi_1, Manfredi_2, Manfredi_3, 56, 57}, plasmonics \cite{ Yan, Ciraci, 52, 53}, and electrons in semiconductors \cite{50, 51} and is, therefore, in the focus of this paper.

The key ingredients of  QHD -- as it is often used in the context of quantum plasmas -- are the ideal Fermi pressure, $P_F$, and the so-called Bohm potential, $V_B$ \cite{Khan}.
%\textcolor{red}{
In QHD, the closed momentum equation is, instead of Eq.~(\ref{TDFT_2}),
\begin{multline}\label{QHD_F_B}
m \frac{\partial}{\partial t} \vec j\left(\vec r, t\right) - n \left(\vec r, t\right) e\vec E (\vec r, t) = \\
-\vec{\nabla} P_F[n(\vec r, t)] -n\left(\vec r, t\right)\vec{\nabla}V_B[n(\vec r, t)],
\end{multline}
%}
%
Manfredi and  Haas \cite{Manfredi_2} derived the Fermi pressure and a  quantum correction in the form of the Bohm potential using a semi-classical Hartree ansatz for the $N$-electron wave functions with identical amplitude 
%and phase 
for all single-electron orbitals \cite{Khan}.
However, in order to reach agreement with the results of the more fundamental kinetic theory in its simplest form -- the random phase approximation (RPA) -- both, the Fermi pressure and the Bohm potential have to be ``corrected'' 
%by multiplying 
by constant pre-factors \cite{Halevi, Yan, akbari_pp15, POP2015}.
Similarly, in plasmonics, the QHD theory is used with one or, sometimes, two fitting parameters corresponding to the prefactors of the Fermi pressure and the Bohm potential, but with an additional exchange correlation potential contribution, which is valid only  in the static case.

Moreover, in the context of quantum plasmas, QHD has often been used beyond the range of applicability of the model and, occasionally, even with
explicitly incorrect expressions. This has led to un-physical predictions and some controversy, for a discussion, see Refs. \cite{Khan, Krishnaswami, Schoof_1, Schoof_2, michta_cpp15}.

However, introducing the above mentioned corrections factors does not solve the problem.
It turned out that these fitting parameters (pre-factors) are not constants but 
differ depending on the characteristic wavelength and frequency of the physical problem. In addition, these coefficients are found to vary with the plasma density and temperature. This results in complicated parametric dependencies of the Bohm potential. 
This means that the range of the values in which correcting parameters can be chosen and the corresponding underlying physical assumptions need to be clarified.

For this reason, in this paper QHD is reconsidered in detail, for both zero temperature and finite temperature, and it is put into the context of well established approaches such as Thomas-Fermi theory and dielectric theory (such as the random phase approximation). Starting from these approaches it becomes more clear what approximations are actually being made and what is the corresponding range of validity of the model.

In Sec.~\ref{s:2}, continuity and momentum equations of the QHD model are  briefly introduced for the finite temperature case.
In Sec.~\ref{s:gradient_app},  the relation of the Bohm potential in the density gradient approximation to the power expansion of the inverse finite temperature RPA polarization function is presented, and
the QHD potentials, i.e., the potential related to the Fermi pressure  and the Bohm potential, are considered in different limiting cases.
This will allow us to reproduce known results and, in part, even to improve them.  
In Sec.~\ref{s:nonlocal_Bohm}, the generalized non-local Bohm potential, based on the exact RPA polarization function, is presented. 
The exchange-correlation potential for the QHD application is discussed  in Sec.~\ref{s:XC}.
The paper is concluded by a discussion of the range of applicability of the QHD model.

% The correct factor $1/9$ for Bohm potential in static case was derived recently on the basis of RPA polarization function \cite{POP2015}. 

%----------------------
\section{QHD equations at finite temperatures}\label{s:2}
The underlying equations of the QHD model can be derived from a field theory, starting from the semi-classical Hamiltonian which, in the absence of a magnetic field, reads \cite{Ying, Yan}

  \begin{multline}\label{Hamiltonia}
H[n(\vec r, t), w(\vec r, t)]=E[n(\vec r, t)]-\int e V_{\rm ext}n(\vec r, t)\mathrm{d}\vec{r} \\
+\int\frac{m_en(\vec r, t)}{2}\left|\vec{\nabla} w\left(\vec r, t\right)\right|^2 \mathrm{d}\vec{r}
+\frac{e^2}{2}\int\frac{n(\vec r, t)n({\vec r}^{\prime}, t)}{\left|\vec r-\vec r^{\prime}\right|}\mathrm{d}\vec{r}\mathrm{d}\vec{r}^{\prime},
 \end{multline} 
where $w$ is the scalar potential determining the velocity field by $\vec{v}=-\vec{\nabla} w$,
$E[n]=E_{\rm id}[n]+E_{\rm xc}[n]$ is the sum of the kinetic and the exchange-correlation energy % density 
functionals, and $V_{\rm ext}$ refers to the external electric potential.   

Using $n(\vec r, t)$ and $m_e w(\vec r, t)$ as canonically conjugate field variables, we employ Hamilton's equations \cite{Lurie}:
\begin{align}\label{H_eq1}
\frac{\delta H[n(\vec r, t), w(\vec r, t)]}{m_e\delta w(\vec r, t)} &=-\frac{\partial n(\vec r, t)}{\partial t} ,
\\
\frac{\delta H[n(\vec r, t), w(\vec r, t)]}{\delta n(\vec r, t)} &=m_e\frac{\partial w(\vec r, t)}{\partial t} ,
\label{H_eq2}
 \end{align}
and obtain the following equations of motion that form the basis of QHD \cite{Ying, Banerjee}:
\begin{align}\label{QHD1}
\frac{\partial}{\partial t} n\left(\vec r, t\right)&=\vec{\nabla}\left[n\left(\vec r, t\right)\vec{\nabla}w\left(\vec r, t\right)\right],\\
m_e \frac{\partial}{\partial t} w\left(\vec r, t\right)&=\frac{\delta E[n]}{\delta n}-e V_{\rm ext}+\nonumber\\
&\phantom{{}=} e^2\int\frac{n({\vec r}^{\prime}, t)}{\left|\vec r-\vec r^{\prime}\right|}\mathrm{d}\vec{r}^{\prime} +
\frac{e^2}{2}m_e\left|\vec{\nabla}w\left(\vec r, t\right)\right|^2.\label{QHD2}
\end{align}
Introducing the potential of the generalized force \cite{michta_cpp15} 
  \begin{equation}\label{chem}
\mu[n(\vec r ,t)]=\frac{\delta E[n(\vec r, t)]}{\delta n(\vec r, t)}+e\varphi(\vec r, t),
 \end{equation} 
 with the definition
 \begin{equation}\label{phi} \varphi(\vec r, t)=e\int\frac{n({\vec r}^{\prime}, t)}{\left|\vec r-\vec r^{\prime}\right|}\mathrm{d}\vec{r}^{\prime}-V_{\rm ext},
 \end{equation}
 and making use of relations $\vec{v}=-\vec{\nabla} w$ and $(\vec v \cdot \vec \nabla)\vec v=\frac{1}{2}\vec\nabla(\vec \nabla w)^2$,  we arrive at the QHD equation in terms of average density $n\left(\vec r, t\right)$, velocity
  $v\left(\vec r, t\right)$, and the generalized force $-\vec{\nabla} \mu \left[\vec r, t\right]$  \cite{Ying, Banerjee, michta_cpp15}
\begin{align}\label{QHD_3}
\frac{\partial}{\partial t} n\left(\vec r, t\right)+\vec{\nabla} \left[n\left(\vec r, t\right)v\left(\vec r, t\right)\right]&=0,
\\
m_e \frac{\partial}{\partial t} \vec{v}\left(\vec r, t\right)+m_e \vec{v}\left(\vec r, t\right)\vec{\nabla} \vec{v}\left(\vec r, t\right)&=-\vec{\nabla} \mu \left(\vec r, t\right).
\label{QHD_4}
 \end{align}

In previous works, Eq.~(\ref{QHD_4}) was obtained for the case of fully degenerate electrons (zero temperature limit). 
%In general, 
Equations~(\ref{QHD_3}) and (\ref{QHD_4}) represent QHD equations in the micro-canonical ensemble, as they were derived from the semi-classical Hamiltonian (\ref{Hamiltonia}) (alternatively, they can be obtained from a semi-classical Lagrangian \cite{Ying}). 

%\textbf{for later reference we should also rewrite the momentum equation for $T=0$ with the Fermi pressure and Bohm potential.
%}

For the extension to finite temperature, we now switch to the grand canonical ensemble. There, these equations have the same form, but $n$ and $w$ must be understood as quantities that are averaged over the grand ensemble \cite{Castro, Zubarev}.
Now we generalize the momentum equation (\ref{QHD_4}) to the finite temperature case 
%noting that in the finite temperature theory 
%it is convenient 
where it is advantageous to use the free energy %density 
functional, $F[n]$, instead of $E[n]$. 
Indeed, in the grand  canonical ensemble we have \cite{Zubarev}

\begin{equation}
\frac{\delta E}{\delta n}=\frac{\delta \Omega}{\delta n},
\label{dE=dW}
\end{equation}
 with the grand potential 
 \begin{equation}
 \Omega [n(\vec r)]=F[n(\vec r)]-\mu_0N,
 \label{Omega}
 \end{equation}
 where $\mu_0 $ is a constant
 defining the chemical potential of the system in thermodynamic equilibrium, and $N=\int n(\vec r) \mathrm{d}\vec{r}$ corresponds to the mean value of the number of particles in the grand canonical ensemble. The derivation 
 of Eq.~(\ref{dE=dW}) is given in the Appendix A. 
 
 It is should be noted that, in Eq.~(\ref{dE=dW}), $E$ is the average value of the energy over a grand canonical ensemble. 
 With this, we obtain for  the potential   of the generalized force at finite temperature:
 \begin{equation}\label{chem_T}
\mu[n(\vec r ,t), T]+\mu_0=\frac{\delta F[n(\vec r, t)]}{\delta n(\vec r, t)}+e\varphi(\vec r, t),
 \end{equation} 
where $F[n]=F_{\rm id}[n]+F_{\rm xc}[n]$ is decomposed into the ideal (non-interacting) part  $F_{\rm id}[n]$, and the exchange correlation part $F_{\rm xc}[n]$.
 In the static long wavelength limit (see below), the generalized force, $-\vec \nabla \mu[n(\vec r ,t), T]$, is related to the pressure $P$ \cite{Chihara}
  \begin{equation}
    \vec \nabla P=n\vec \nabla \frac{\delta F[n(\vec r, t)]}{\delta n(\vec r, t)},
  \end{equation}
which provides the link with the standard fluid theory, cf. Eq.~(\ref{TDFT_2}).

 \section{Derivation of the Bohm potential and Fermi pressure}\label{s:gradient_app}
\subsection{General Expressions}
In order to derive a potential related to the  Fermi pressure and the Bohm potential, we neglect the exchange-correlation term, $F_{\rm xc}[n]$, and turn to the local density approximation (LDA) with non-locality taken into account by the first order gradient correction 
to the noninteracting free energy %density 
functional \cite{Mermin, POP2015, Note}:
\begin{align}
F_{\rm id}[n] &= F_0[n] + \int \mathrm{d}\vec{r}\,a_{2}\left[n\right] \mid \vec{\nabla} n(\vec{r})\mid^{2},
\label{T}
\end{align}
where the  free energy $F_{0}$ is defined via the free energy density  
   \begin{equation}\label{T_0}
F_{0}[n]=\int f_0[n] \mathrm{d}\vec r,
 \end{equation} 
and the functional $a_2[n]$ still remains to be found.
Substituting Eq.~(\ref{T}) into (\ref{chem_T}) we have:
  \begin{multline}\label{chem_T2}
\mu_{\rm id}[n(\vec r ,t), T]+\mu_0=\frac{\partial f_0[n]}{\partial n}+\frac{\partial a_{2}\left[n\right]}{\partial n}\mid \vec{\nabla} n(\vec{r})\mid^{2}-\\
2{\vec \nabla} \left[a_{2}[n] \vec{\nabla} n(\vec{r})\right] +e\varphi(\vec r, t).
 \end{multline} 
 
Now we show that it is possible to connect the QHD Bohm potential with the expansion of the inverse polarization function and, thereby, to obtain $a_2$.
For the static case, $\omega=0$, this was done by Perrot \cite{Perrot}.
Here we generalize his results to the dynamic case.
We consider small localized variations of the potential, $\delta \varphi$, density,  $\delta n$, velocity in terms of the scalar potential, $\delta w$, and of the chemical potential, $\delta \mu_{\rm id}$. 
From Eq.~(\ref{QHD1}), taking the equilibrium density distribution as uniform ($n_0={\rm const}$) and assuming $w_0=0$, we obtain
    \begin{equation}
%    \label{d_n}
%-i\omega \delta {\tilde n} +k^2n_0 { \delta {\tilde w}}=0,
\delta {\tilde w}=\frac{i\omega}{k^2 n_0} {\delta {\tilde n}},
\label{eq:deltaw_deltan}
 \end{equation} 
where $\delta \tilde n$, and $\delta \tilde w$ denote the Fourier transforms of $\delta n$ and $\delta w$, respectively.
%From (\ref{d_n}), we find $\delta {\tilde  w}=\frac{i\omega}{k^2 n_0} {\delta {\tilde n}}$. 
%
The resulting  variation of $\mu_{\rm id}$, as defined by Eq.~(\ref{chem_T2}), has the form \cite{Perrot}
\begin{equation}\label{chem_T3}
\delta \mu_{\rm id}=\left(\frac{\partial^2 f_0[n]}{\partial n^2}\Big|_{\rm n=n_0}-2a_{2}[n_0] \Delta \right)\delta n  +e\delta \varphi.
 \end{equation} 
Thus, the linearized momentum equation (\ref{QHD_4}) for $\tilde w=\delta \tilde w$ and, taking into account Eq.~(\ref{chem_T3}), is written as:
     \begin{multline}\label{lin_m}
 -i\omega m_e\delta \tilde w=e{ \delta \tilde \varphi} +\\
\left[\left.\frac{\partial^2f_0[n]}{\partial n^2} \right|_{\rm n=n_0} +2a_2[n_0]k^2 \right] { \delta \tilde n}.
 \end{multline} 
Finally, making use of Eq.~(\ref{eq:deltaw_deltan}),
%   $\delta {\tilde  w}=\frac{i\omega}{k^2 n_0} {\delta {\tilde n}}$ into Eq.~(\ref{lin_m}), 
we find 
  \begin{equation}\label{dV}
{e\delta {\tilde \varphi}}=-\left[\left.\frac{\partial^2f_0[n]}{\partial n^2} \right|_{\rm n=n_0} +2a_2[n_0]k^2 -\frac{\omega^2}{k^2}\frac{m_e}{n_0}\right] \delta {\tilde n}.
 \end{equation} 

This is an important result that relates the density perturbation to the perturbation of the external potential. As we have assumed weak perturbations, we can make use of the results of linear response theory and identify in Eq.~(\ref{dV})  the inverse polarization function, $\Pi^{-1} \equiv e \delta {\tilde \varphi}/\delta {\tilde n}$:
    \begin{equation}\label{Pol_2}
\frac{1}{\Pi\left(k, \omega\right)}=-\left[\frac{\partial^2f_0[n]}{\partial n^2}\Big|_{\rm n=n_0}  +2a_2[n_0]k^2 -\frac{\omega^2}{k^2}\frac{m_e}{n_0} \right].
 \end{equation} 
 
 \subsection{RPA result for the Bohm potential and Fermi pressure}
Equation (\ref{Pol_2}) gives us the opportunity to express the r.h.s. of Eq.~(\ref{Pol_2}) systematically via linear response theory. The lowest order many-body approximation for $\Pi$ is the random phase approximation (RPA), which reads in equilibrium, for arbitrary temperature \cite{Arista} 
\begin{equation}\label{Pi}
\Pi_{\rm RPA}(k,\omega)=-\frac{k^2\chi_{0}^{2}}{16 \pi e^2z^3}\left[g(u+z)-g(u-z)\right],
\end{equation}
where we introduced dimensionless frequency and wave number,  $u=\omega/(kv_F)$, $z=k/(2k_F)$, and defined $\chi_{0}^{2}=(\pi k_F a_B)^{-1}\simeq r_s/6.03$, where $a_B$ is the Bohr radius, $r_s \equiv a/a_B$, with $a$ being the mean interparticle distance, and we also introduced the Fermi wave number and the plasma frequency, $k_F=(3\pi^2n)^{1/3}$, $\omega_{p}^{2}=4\pi ne^2/m_e$. Finally, we defined
\begin{equation}\label{g}
g(x)=-g(-x)=\int \! \frac{y\,\mathrm{d}y}{\exp(y^2/\theta-\eta)+1}\ln\left|\frac{x+y}{x-y}\right|.
\end{equation}

With these explicit expressions the RPA polarization can be studied in detail. Particularly simple expressions exist for various limiting cases with respect to frequency and wave number.
In the limiting cases of large or small values of $z$, the inverse of the real part of the RPA polarization function has the following expansion \cite{Arista, Wang2000}:
\begin{multline}\label{Pi_expansion}
\frac{1}{2\Pi_{\rm RPA}(z,u)}\simeq {\tilde a_0}+{\tilde a_2}(2k_F)^2~z^2+{\tilde a_4}(2k_F)^4~z^4+...+c u^2\\
\simeq {\tilde a_0}+{\tilde a_2}k^2+{\tilde a_4}k^4+...+\frac{1}{2}\frac{\omega^2}{k^2}\frac{m_e}{n_0}.
\end{multline}
In particular, we obtain the long-wavelength limit of the inverse polarization function as 
\begin{equation}
\frac{1}{\Pi^0(\omega)} \equiv \lim_{k\to 0}\frac{1}{\Pi_{\rm RPA}(z,u)} 
= \frac{\omega^2}{k^2}\frac{m_e}{n_0}.
\label{eq:pi0}
\end{equation}
 Comparing Eqs.~(\ref{Pol_2}) and (\ref{Pi_expansion}), we see that 
  \begin{align}\label{term_1}
{\tilde a_{0}}[n_0]&=-\left. \frac{1}{2}\frac{\partial^2f_0[n]}{\partial n^2}\right|_{\rm n=n_0},
\\
 {\tilde a}_2[n_0]&=-a_2[n_0].
\label{term_2}
\end{align}
 The convergence of the expansion (\ref{Pi_expansion}) is discussed in Appendix B for the static case and $k<2k_F$.

We stress that the coefficients  $ {\tilde{a}_{0}}$ and $ {a_{2}}$ depend on the considered  limits for $k$ and $\omega$ which directly affects the value of the Bohm potential, as will be shown below. 
%\textcolor{red}{
Equations~(\ref{term_1}) and (\ref{term_2}) are closely related to the  \textit{stiffness} theorem connecting the perturbation of an equilibrium state (of the energy, in the case of the ground state) due to an applied external field with an inverse linear response function.  The proof of this theorem was given in Ref.~\cite{Giuliani} for the ground state and can be easily extended to finite temperatures.
%}   

Once the coefficients  $ {\tilde{a}_{0}}[n_0]$ and $ {a_{2}}[n_0]$ are defined,  we allow the equilibrium density to vary in space and time, $n_0 \to n(\vec r, t)$, according to the standard concept of LDA. 
The Bohm potential in the local density approximation, for QHD applications can be obtained 
for any degeneracy parameter from the second term on the right hand side of Eq.~(\ref{T}):
\begin{multline}\label{Bohm_pot}
V_{B}= \frac{\delta}{\delta n}\int \mathrm{d}\vec{r}\,a_{2}\left[n\right] \mid \vec{\nabla} n(\vec{r})\mid^{2}\\
=\left|\vec{\nabla} n\right|^2\frac{\partial a_2[n]}{\partial n}-2\left(a_2[n]\vec{\nabla}^2 n+\vec{\nabla} n \vec{\nabla} a_2[n] \right).
 \end{multline} 
The coefficients  $ {\tilde{a}_{0}}[n]$ and $ {a_{2}}[n]$ are expressed in terms of the density $n(\vec r)$ and  the dimensionless chemical potential, $\eta=\beta\mu$, where we introduced the inverse temperature, $\beta=1/(k_BT)$.
%is the inverse value of the thermal energy.

Using the local density approximation, we obtain $n(\vec r)=\frac{\sqrt{2} m^{3/2}}{\pi^2\beta^{3/2}\hbar^3}I_{1/2}[\eta(\vec r)]$. The following partial derivatives are 
 needed to find the Bohm potential (\ref{Bohm_pot}), 
% and can be written as:
\begin{align}\label{d_eta}
\frac{\partial \eta}{\partial n}=C\frac{1}{I_{-1/2}(\eta)},
\\
 \vec \nabla \eta=C\frac{\vec{\nabla} n(\vec r)}{I_{-1/2}(\eta)},
\label{d_eta2}
 \end{align} 
where $C=\frac{2\pi^2\beta^{3/2}\hbar^3}{\sqrt{2} m^{3/2}}=\frac{4}{3n}\theta^{-3/2}$.
 As is shown below, at $\theta=k_BT/E_{F}\sim T\times n^{-2/3} \to 0$, we have $a_2[n]=\gamma \hbar^2/(8mn)$, and the Bohm potential has the following form:
 \begin{equation}\label{Bohm_pot2}
V_{B}(\omega,k) = \gamma(\omega,k) \frac{\hbar^2}{8m}\left(\left|\frac{\vec{\nabla} n}{n}\right|^2-2\frac{\vec{\nabla}^2n}{n}\right),
 \end{equation} 
where the coefficient $\gamma$ sensitively depends on the considered values of the wave number and frequency.

On the other hand, the Fermi pressure is proportional to the functional derivative of the Thomas-Fermi free energy  \cite{michta_cpp15}, which,  at $\theta\to 0$, can be written as:
 \begin{equation}\label{Fermi_pres}
\frac{\delta F_{0}[n]}{\delta n}=\frac{\partial f_{0}[n]}{\partial n}=-\int 2{\tilde a_0\left([n],\theta \to 0\right)}\mathrm{d}n=\bar{\alpha} E_F,
 \end{equation} 
 where the coefficient $\bar{\alpha}$ depends on the considered limit on 
the \xyplane.
 Previously, this coefficient was chosen by adjusting the QHD result for the longitudinal plasmon dispersion to the RPA prediction \cite{Jones, Halevi} for zero temperature [note that $F\to E$, at $\theta\to 0$]. 

Now, we will use  Eq.~(\ref{term_1}), to analyze different limits for the coefficients  $ {\tilde{a}_{0}}[n]$ and $ {a_{2}}[n]=-\tilde{a}_2$, for different frequency-wavenumber ranges.\\

\subsection{Low frequency and long wavelength limit,   $\omega\ll \hbar k^2/2m, k\ll 2k_F$}
The results given for the static case remain valid also for low frequencies, i.e. $u\ll 1$ or $\omega\ll kv_F$. %\textcolor{red}{
This regime is relevant for a variety of physical processes such as the screening of a test charge  and the dispersion of low-frequency waves (e.g. ion-acoustic waves) in a plasma. In a variety of publications e.g. the topic of screening of an ion in a quantum plasma were treated incorrectly using the Bohm potential for the high-frequency case rather than for the static case. This has lead to the unphysical prediction of ion-ion attraction in an equilibrium quantum plasma, for a discussion, see Refs.~\cite{Schoof_1, POP2015}.
%}
 %
 \begin{figure}[t]
\includegraphics[width=0.39\textwidth]{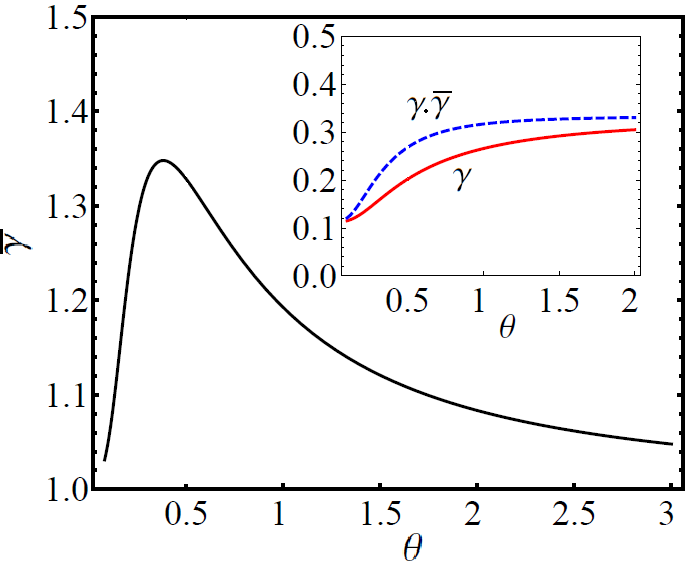}
\caption{The values of the factors $\bar{\gamma}$, $\gamma$, and $\gamma \bar{\gamma}$ in the  finite temperature Bohm potential, Eq.~(\ref{Bohm_pot_finite_T}), for the long-wavelength case.}
\label{fig:gamma}
\end{figure}
To obtain the Fermi pressure and Bohm potential in the present limiting case, we calculate the coefficients  $ {\tilde a_{0}}[n]$ and $ {a_{2}}[n]$ and obtain~\cite{POP2015}:
\begin{align}\label{a0}
{\tilde a_0}\left[n\right]&=-\frac{\pi e^2}{2k_F^2\chi_0^2 H_1(\eta)},
\\
a_2\left[n\right] &=-\frac{\hbar^2 H_2(\eta)}{72m_e n H_1^2(\eta)},
\label{a2}
\end{align}
with 

$H_1(\eta)= \frac{\sqrt{\theta[n]}}{2} \, I_{-1/2}(\eta)$, $\quad H_2(\eta)=\frac{1}{2\sqrt{\theta[n]}} \,I_{-3/2}(\eta)$.
In the low-temperature limit, $\theta\rightarrow 0$, we have $H_1(\eta)\simeq 1$ and $H_2\simeq -1$, and Eq.~(\ref{a2}) gives the asymptotic results $a_2\to \hbar^2/(72mn)$ and $\gamma = 1/9$ for the coefficient in front of the Bohm potential in Eq.~(\ref{Bohm_pot2}) \cite{POP2015, Perrot, Kompaneets, Kirzhnitz}.

Using the relations (\ref{d_eta}) and (\ref{d_eta2}), the finite-temperature generalization of the Bohm potential can be easily found by substituting Eq.~(\ref{a2}) into Eq.~(\ref{Bohm_pot}) and taking into account the dependence $\theta [n] \sim n^{-2/3}$.
Furthermore, for the static case ($\omega=0$), we have $f_0 \equiv f_{\rm TF}$, where $f_{\rm TF}$ is the Thomas-Fermi free energy density:
\begin{equation}\label{TF}
f_{\rm TF}\left([n],\theta\right)=\frac{\sqrt{2}m^{3/2}}{\hbar^3\pi^2\beta^{5/2}}\left(\eta I_{1/2}(\eta)-\frac{2}{3}I_{3/2}(\eta) \right),
 \end{equation} 
 where $I_{\nu}$ is the Fermi integral of order $\nu$.  
  Using Eq.~(\ref{TF}) Perrot \cite{Perrot} showed that the second order partial derivative of the Thomas-Fermi density of states, $-\frac{1}{2}\frac{\partial^2f_{TF}[n]}{\partial n^2}$, in the static long wave length limit ($k\ll 2k_{F}$) 
exactly coincides with ${\tilde a_{0}}[n]$ given by Eq.~(\ref{a0}).  
Finally, the functional derivative of the Thomas-Fermi term yields:
\begin{equation}\label{TF_pres}
\frac{\delta F_{\rm TF}}{\delta n}=\frac{\eta}{\beta}.
% \frac{\sqrt{2}m^{3/2}}{\hbar^3\pi^2\beta^{5/2}}\times \frac{1}{2}\eta I_{1/2}(\eta)\frac{\partial \eta}{\partial n},
 \end{equation} 
 At zero temperature, $\eta=\beta E_F$, leading to $\delta T_{\rm TF}/\delta n=E_F$, in agreement with the result of Ref.~\cite{michta_cpp15}.
 
 By regrouping terms we can rewrite the Bohm potential (\ref{Bohm_pot}) in the following form:
  \begin{equation}\label{V1+V2}
V_{B}=V_1+V_2,
 \end{equation} 
 where 
   \begin{equation}\label{V1}
V_1=-2a_2[n] \vec{\nabla}^2n,
 \end{equation} 
 and
   \begin{align}\label{V2}
V_2&=\left|\vec{\nabla} n\right|^2\frac{\partial a_2[n]}{\partial n}-2\vec{\nabla} n\frac{\partial a_2[n]}{\partial r} \notag \\
&=\left|\vec{\nabla} n\right|^2\frac{\partial \eta}{\partial n}\frac{\partial a_2[n]}{\partial \eta}-2\vec{\nabla} n\frac{\partial \eta}{\partial r} \frac{\partial a_2[n]}{\partial \eta} \notag\\
&=-\left|\vec{\nabla} n\right|^2\frac{C}{I_{-1/2}(\eta)}\frac{\partial a_2[n]}{\partial \eta},
 \end{align} 
and the last line of Eq.~(\ref{V2}) was obtained using relations (\ref{d_eta}) and (\ref{d_eta2}).
 In order to analyze finite temperature effects we introduce the coefficient:
 \begin{equation}\label{gamma_t}
\gamma[n]=a_2\left[n\right] \left(\frac{\hbar^2}{8m_en}\right)^{-1}=-\frac{2}{9}\frac{I_{-3/2}(\eta)}{I_{-1/2}^2(\eta)}\theta^{-3/2},
 \end{equation} 
  similar to the zero-temperature case, and use it in $V_1$ to obtain
     \begin{equation}\label{V1_gamma}
V_1=\gamma \frac{\hbar^2}{8m_e}\left(-2\frac{\vec{\nabla}^2n}{n}\right).
 \end{equation} 
  By taking into account that $C=\frac{4}{3n}\theta^{-3/2}$, we find:
     \begin{equation}\label{V2_gamma}
V_2=-\gamma  \frac{\hbar^2}{8m_e} \frac{2I_{-1/2}(\eta)I_{1/2}(\eta)}{I_{-3/2}(\eta)}\frac{\partial a_2[n]}{\partial \eta}\times\frac{\left|\vec{\nabla} n\right|^2}{n^2}.
 \end{equation} 
 
 Finally, we introduce the coefficient:
 \begin{equation}\label{bar_gamma}
\bar{\gamma}[n]=-\frac{2I_{1/2}(\eta)I_{-1/2}(\eta)}{I_{-3/2}(\eta)}\frac{\partial}{\partial \eta}\left(\frac{I_{-3/2}(\eta)}{I_{-1/2}^2(\eta)}\right),
 \end{equation} 
 and substitute Eqs.~(\ref{V1_gamma}) and (\ref{V2_gamma}) into Eq.~(\ref{V1+V2}). This yields the following expression for the finite-temperature Bohm potential in the static long wavelength limit:
   \begin{equation}\label{Bohm_pot_finite_T}
V_{B}(\theta)=\gamma(\theta) \frac{\hbar^2}{8m}\left(\bar{\gamma}(\theta)\left|\frac{\vec{\nabla} n}{n}\right|^2-2\frac{\vec{\nabla}^2n}{n}\right).
 \end{equation} 

 The correction coefficient $\bar{\gamma}$ of the first term in the Bohm potential for the finite-temperature case (\ref{Bohm_pot_finite_T}), and the dependence of $\gamma$ and $\gamma\bar{\gamma}$ on $\theta$ are presented in Fig.~\ref{fig:gamma}.
 In the limit of fully degenerate electrons, $\theta\to 0$, as well as in the classical limit, $\theta\gg1$, the correction coefficient approaches unity, $\bar{\gamma}\to 1$.  
 This correction coefficient $\bar{\gamma}$ is important for a partially degenerate plasma, around $\theta \sim 0.5$, as can be seen from Fig.~\ref{fig:gamma}.
  
  Recently, Haas and Mahmood \cite{Haas} considered the finite-temperature Bohm potential by analyzing ion-acoustic waves on the basis of  linearized QHD equations and comparing them with the RPA result. 
  %of RPA for a linear ion-acoustic wave dispersion.
  They correctly derived the coefficient $\gamma$ in Eq.~(\ref{gamma_t}), but missed the correction coefficient $\bar{\gamma}$ in Eq.~(\ref{bar_gamma}). 
  Note that in Eq.~(\ref{Bohm_pot_finite_T}), the second term of the Bohm potential is proportional to $n_1/n_0$, whereas the first term $\sim (n_0/n_1)^2$, where $n_1$ is a small density perturbation (i.e., $n_0/n_1\ll1$).
  As a result, in the linear approximation, which was  considered by  Haas and Mahmood \cite{Haas},  the information about the first term of 
  the Bohm potential and  the coefficient $\bar{\gamma}$ is lost. 

%---------------------------
 \subsection{Short wavelength limit, $k\gg 2k_F$ at low frequencies, $\omega\ll \hbar k^2/2m$}
 For the degenerate electron gas, Jones and Yang \cite{Jones} showed that, in the short wavelength limit, the first order gradient correction term of the non-interacting kinetic energy has the form of the von Weizs{\"a}cker gradient correction 
 with $a_2\left[n\right] =\frac{\hbar^2}{8m_e n}$, which gives  $\gamma=1$ for the Bohm potential. 
 This result is important as the von Weizs{\"a}cker gradient correction correctly reproduces  Kato's cusp condition for the electron distribution close to a test charge (core) \cite{Plumper}.
 Knowledge of the change of the coefficients determining the Fermi pressure and Bohm  potential at the transition from the long wavelength limit to the short wavelength
 case can be useful to analyze the QHD results in plasmonics \cite{Ciraci, Pitarke}, quantum plasmas \cite{Haas_2000} and the application of orbital--free DFT to dense plasmas (warm dense matter) \cite{Daligault, Starrett, Karasiev}.
 
 In the low-frequency short wavelength limit, $u\ll z$, $z\gg 1$,  we use the following expansion of the function $\Delta g=g(u+z)-g(u-z)$ \cite{Arista}:
  \begin{align}\label{dg1}
  \Delta g &\simeq 2g(z)+u^2g^{\prime \prime}(z),
  \end{align}
 where
 \begin{align}\label{delta_g_2_z>>1}
g(z) &=\frac{2}{3z}+\frac{I_{3/2}\theta^{5/2}}{3z^3}.
\end{align}
From Eqs.~(\ref{dg1}) and (\ref{delta_g_2_z>>1}) we obtain
  \begin{align}\label{Dg1}
  \frac{z}{\Delta g} &\simeq -\frac{3}{8}I_{3/2}(\eta)\theta^{5/2}+\frac{3z^2}{4}-\frac{3u^2}{4}+...,
  \end{align}
which can be substituted into the formula for the inverse polarization function, with the result
    \begin{align}\label{Pol_4}
  \frac{1}{\Pi (k,\omega)} & = -\frac{4\pi e^2}{\chi_0^2 k_F^2}\frac{z}{\Delta g}.
  \end{align}
From this,  we obtain  the coefficients ${\tilde a}_0[n]$ and $a_2[n]$ in the short wavelength case:
 \begin{align}\label{a0_z<1}
{\tilde a_0}\left[n\right]&=-\frac{4\pi e^2}{k_F^2\chi_0^2}\times \frac{3}{16}I_{3/2}(\eta)\theta^{5/2}[n],
\end{align}
 \begin{align}\label{a2_z>1}
a_2\left[n\right] &=\frac{\hbar^2}{8m_e n}.
\end{align}
   \begin{figure}[t]
\includegraphics[width=0.38\textwidth]{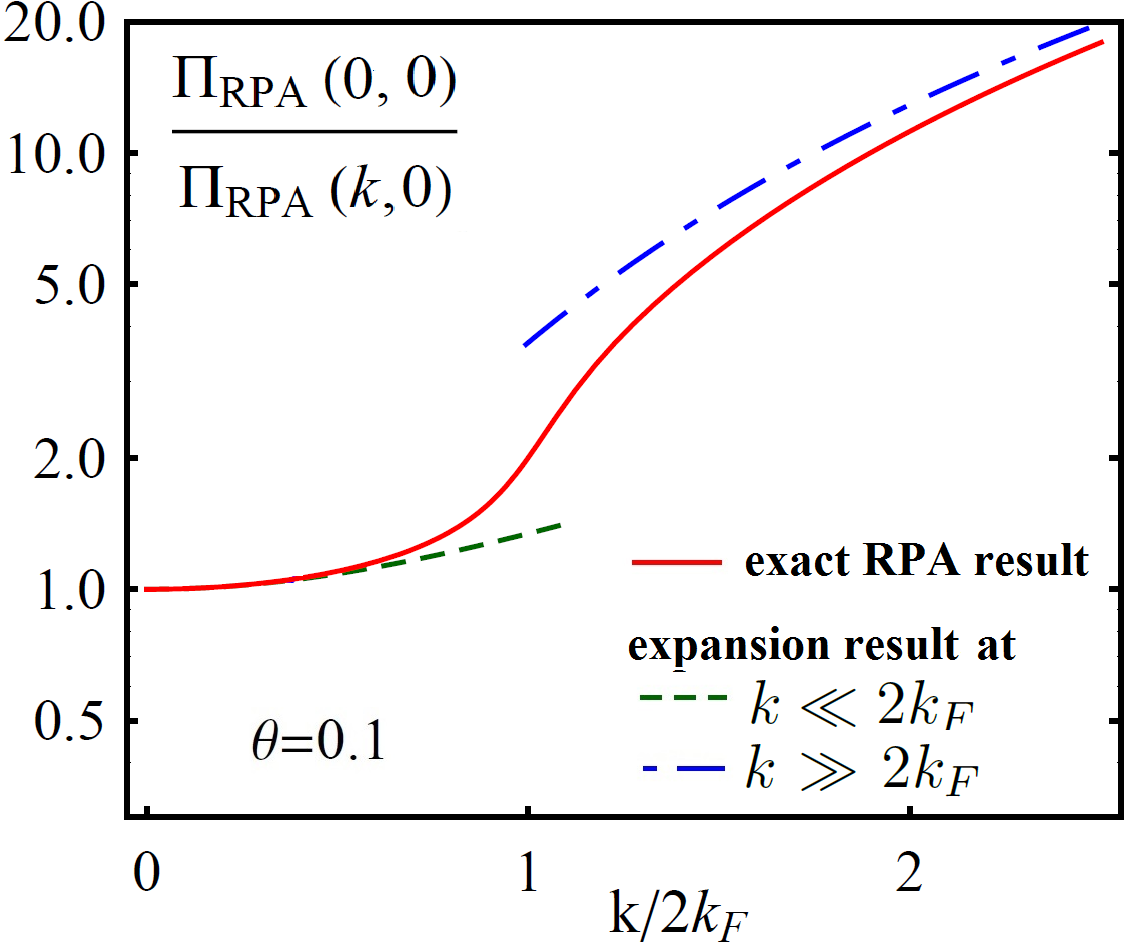}
\caption{The values of the quantity $\Pi_{\rm RPA}(0,0)/\Pi_{\rm RPA}(k,0)$ at $\theta=0.1$.}
\label{fig:Pi_static}
\end{figure}
 
From Eq.~(\ref{a2_z>1}) we see that the coefficient in front of the Bohm potential in Eq.~(\ref{Bohm_pot2}) becomes $\gamma = 1$ \cite{Jones}, and it is interesting to note that, in the short wavelength limit, the Bohm potential is independent of temperature. 
Consider now the Fermi pressure term. 
At small temperatures, $\theta\ll1$, using Eq.~(\ref{a0_z<1}) for the  functional derivative of $F_0[n]$ yields
\begin{equation}\label{TF_pres_theta<1_u<z}
\frac{\delta F_{0}[n]}{\delta n}=\frac{3}{5}E_F+\frac{3\pi^2}{8}E_F\theta^2,
 \end{equation} 
 where the first term is the result for the ground state and the second term is the finite-temperature correction.
Thereby, the transition from the long wavelength to the short wavelength limit leads to a change of the factor $\bar{\alpha}$, from $1$ to  $3/5$.
Note that, in the short wavelength limit, or for large values of the density gradient the von Weizs\"acker gradient correction is the leading term in the non-interacting free energy functional of electrons \cite{Plumper}.

 In Fig.~\ref{fig:Pi_static} the results of the expansions of the inverse RPA polarization function in the limits of long, $k/2k_F\ll1$, and short wavelengths, $k/2k_{F}\gg1$, are compared with the exact RPA result.
 It can be concluded that, in the case of the uniform electron gas,  the long-wavelength limit result is applicable up to $k\simeq k_F$.
 \subsection{High frequency limit, $\omega\gg \hbar k^2/2m $} 
 %
%\textcolor{red}{
The present limiting case is important for a variety of physical situations, such as for the description of the optical (Lengmuir) plasmon of the electrons as well as for the plasma response to a high-frequency electromagnetic field with frequencies exceeding the plasma frequency.
%The correct determination of the pre-factors of the Bohm potential and Fermi pressure  
%in the high frequency case is highly important 
 %}
 
To obtain the correct pre-factors of the Bohm potential and Fermi pressure in the present high-frequency limit, $u\gg z$, we use the formulas \cite{Arista}
\begin{align}\label{delta_g}
g(u+z)-g(u-z) &=2zg^{\prime}(u)+\frac{z^3}{3}g^{\prime \prime \prime}(u),
\\
g(u) &=\frac{2}{3u}+\frac{I_{3/2}\theta^{5/2}}{3u^3},
\end{align}
and obtain
  \begin{align}\label{Dg2}
  \frac{z}{\Delta g} &\simeq -\frac{9}{8}I_{3/2}(\eta)\theta^{5/2}+\frac{3z^2}{4}-\frac{3u^2}{4}+...
  \end{align}
The coefficients $ {\tilde a_{0}}[n]$ and $ {a_{2}}[n]$ are obtained from  Eqs.~(\ref{Dg2}) and (\ref{Pol_4})
\begin{align}\label{a0_case2}
{\tilde a_0}\left[n\right]&=-\frac{4\pi e^2}{k_F^2\chi_0^2}\times \frac{9}{16}I_{3/2}(\eta)\theta^{5/2}[n],
\\
a_2\left[n\right] &=\frac{\hbar^2}{8m_e n}.
\label{a2_case2}
\end{align}

Equation~(\ref{a2_case2}) shows that the coefficient in front of the Bohm potential equals $\gamma=1$. In the high-frequency limit
the coefficient $a_2[n]$ does not depend on temperature, which means that at finite temperature the Bohm potential is given by Eq.~(\ref{Bohm_pot2}).
This is  explained by the fact that, at sufficiently high frequency of the perturbing electric field, the back and forth movement of the electrons is not affected by their thermal motion. 

From Eq.~(\ref{a0_case2}), the high-frequency analogue of the Thomas-Fermi pressure can be obtained using $-\int 2{\tilde a_0[n]}\mathrm{d}n$.
In the zero-temperature limit, taking into account the relation $I_{3/2}(\theta\to0)=\frac{2}{5}\theta^{-5/2}$ leads to the following expression for the functional derivative of $F_{0}[n]$, at high frequency:
\begin{equation}\label{TF_pres_theta=0}
\frac{\delta F_{0}[n]}{\delta n}=\frac{9}{5}E_F.
 \end{equation} 
Equation~(\ref{TF_pres_theta=0}) indicates that,  in the high-frequency limit, the QHD equations contain the coefficient $\bar{\alpha}=9/5$.
We note that, in previous works \cite{Yan, Halevi},  this constant coefficient was artificially added in order to reach agreement between the results of the QHD and the  RPA expression for the plasmon dispersion relation. 

A finite-temperature correction to Eq.~(\ref{TF_pres_theta=0}) can be obtained by expanding $I_{3/2}(\eta)$ around $\theta=0$,
\begin{equation}\label{TF_pres_theta<1}
\frac{\delta F_{0}[n]}{\delta n}=\frac{9}{5}E_F+\frac{9\pi^2}{8}E_F\theta^2.
 \end{equation} 
  \begin{figure}[t]
\includegraphics[width=0.45\textwidth]{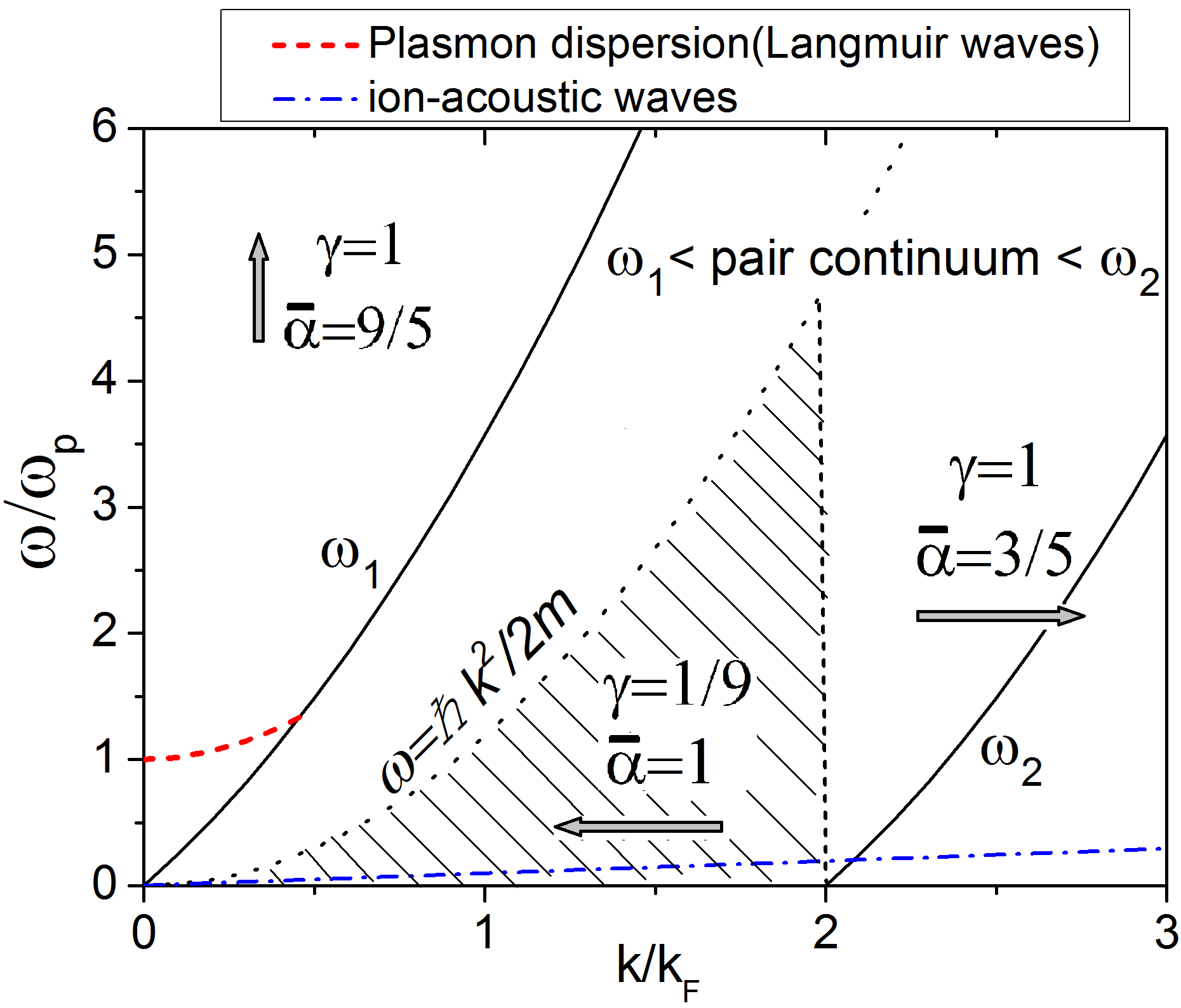}
\caption{The values of the factors $\gamma$ in Bohm potential Eq.(\ref{Bohm_pot2}) and $\bar{\alpha}$ in Eq.(\ref{Fermi_pres}) for different considered cases on $(k, \omega)$ plane at $\theta\ll1$.
The area between curves $\omega_1$ and $\omega_2$ corresponds to the pair continuum. Dashed line is the plasmon dispersion $\omega=\omega_p^2+(3/5)k^2 v_F^2+(1/4)\hbar^2k^4/m_e^2$ and dash-dot line is frequency of  the ion-acoustic waves multiplied by $m_p/m_e$.  
Dashed area corresponds to the static long wave length limit. Arrows indicate that the factors $\gamma$ and $\bar{\alpha}$ were obtained for different limits (see Tables \ref{t:a0} and \ref{t:a2}).}
\label{fig:wk}
\end{figure}
The different values of the factors $\gamma$ and $\bar{\alpha}$ in the \xyplane ~are summarized in Fig. {\ref{fig:wk}}.
Furthermore, the information about the obtained coefficients $a_2 [n], \gamma$ and $\tilde a_0 [n]$,~$\bar{\alpha}$ is listed in tables \ref{t:a2} and \ref{t:a0}. 
\begin{table}[h]\footnotesize
\caption{The prefactor of the Bohm potential $\gamma$ [cf. Eq.~(\ref{Bohm_pot_finite_T})] and the expansion coefficient $a_2 [n]$  in different limits}
\begin{center}
\begin{tabular}{|l|*{2}{c|}r|}
\hline\noalign{\smallskip}
            &  \normalsize{$a_2 [n]$}& \normalsize{$\gamma$} \\
\hline\noalign{\smallskip}
\normalsize{$ \omega\ll \hbar k^2/2m_e$}	   & \large{$-\frac{\hbar^2 H_2(\eta)}{72m_e n H_1^2(\eta)}$} & \normalsize{$1/9$} \\
\normalsize{$k\ll2k_{F}$} & $ $& \\
\hline\noalign{\smallskip}
\normalsize{$  \omega\ll \hbar k^2/2m_e$}		   & \large{$\frac{\hbar^2}{8m_e n}$} & \normalsize{$1$} \\
\normalsize{$k\gg 2k_{F}$} & $ $& \\
\hline\noalign{\smallskip}
\normalsize{$\omega\gg\frac{\hbar k^2}{2m_e}$}	   & \large{$\frac{\hbar^2}{8m_e n}$} & \normalsize{$1$} \\
\noalign{\smallskip}
\hline\noalign{\smallskip}
\end{tabular}
\end{center}
\label{t:a2}
\end{table}
\begin{table}[h]\footnotesize
\caption{The prefactor $\bar{\alpha}$ of the Fermi pressure [cf. Eq.~(\ref{Fermi_pres})] and the expansion coefficient $\tilde a_0 [n]$ in different limits.}
\begin{center}
\begin{tabular}{|l|*{2}{c|}r|}
\hline\noalign{\smallskip}
            &  \normalsize{$\tilde a_0 [n]$}& \normalsize{$\bar{\alpha}$} \\
\hline\noalign{\smallskip}
\normalsize{$ \omega\ll \frac{\hbar k^2}{2m_e}$}	   & \large{$-\frac{\pi e^2}{2k_F^2\chi_0^2 H_1(\eta)}$} & \normalsize{$1$} \\
\normalsize{$k\ll2k_{F}$} & $ $& \\
\hline\noalign{\smallskip}
\normalsize{$  \omega\ll \frac{\hbar k^2}{2m_e}$ }		   & \large{$-\frac{4\pi e^2}{k_F^2\chi_0^2}\times \frac{3}{16}$}\normalsize{$I_{3/2}(\eta)\theta^{5/2}$} & \large{$\frac{3}{5}$} \\
\normalsize{$k\gg 2k_{F}$} & $ $& \\
\hline\noalign{\smallskip}
\normalsize{$\omega\gg\frac{\hbar k^2}{2m_e}$}	   & \large{$-\frac{4\pi e^2}{k_F^2\chi_0^2}\times \frac{9}{16}$}\normalsize{$I_{3/2}(\eta)\theta^{5/2}$} & \large{$\frac{9}{5}$} \\
\noalign{\smallskip}
\hline\noalign{\smallskip}
\end{tabular}
\end{center}
\label{t:a0}
\end{table}

Using the high-frequency result for $\tilde{a}_0$ and the Bohm potential, one can derive the well-known plasmon dispersion from the continuity and momentum equations:
 \begin{equation}
  \omega^2(k) = \omega_p^2-2\tilde{a}_0[n_0]\frac{n_0}{m_e}\times k^2+\frac{\hbar^2k^4}{4m_e^2},
 \end{equation}
which, at $\theta\ll 1$, approaches the form
 \begin{equation}
  \omega^2(k) = \omega_p^2+\frac{3}{5} v_F^2 k^2+\frac{\hbar^2k^4}{4m_e^2},
 \end{equation}
where we took into account that $2\tilde{a}_0[n_0] n_0/m_e \to -\frac{3}{5} v_F^2$, at $\theta \to 0$, and that, in the high-frequency limit, $a_2[n]=\hbar^2/8m_en$.   

On the other hand, in the limit $\theta\gg1$, taking into account that $\tilde{a}_0[n_0] \to -\frac{3}{2}\frac{k_BT}{n_0}$, leads to the dispersion relation
 \begin{equation}
  \omega^2(k) = \omega_p^2+3 v_{\rm th}^2 k^2+\frac{\hbar^2k^4}{4m_e^2},
 \end{equation}
where $v_{\rm th}=\sqrt{\frac{k_BT}{m_e}}$ is the thermal velocity of the electrons.

Note that the linearized QHD equations correctly reproduce the dynamic RPA polarization function in the long wavelength limit, $\Pi^0(\omega)$, Eq.~(\ref{eq:pi0}), without any additional terms related to the fermionic pressure.
%---------------------------------------
\section{Generalized non-local Bohm potential in linear response}\label{s:nonlocal_Bohm}
In the previous section, the Bohm potential of quantum hydrodynamics was derived on the basis of the  RPA polarization function, using the expansion of the latter in powers of the wavenumber. This has allowed us to improve the QHD model, depending on the wavenumber and frequency range in three relevant cases, by   
%\textcolor{red}{
involving the dynamic LDA and the gradient correction, on the basis of the RPA
%}.

On the other hand, QHD can be used to independently obtain the response function of (uncorrelated) electrons, $\Pi^{\rm id}_{\rm QHD}(k,\omega)$, for arbitrary frequencies and wavenumbers. In particular, at large wavenumbers the previous power expansion is not applicable. It is, therefore, instructive to inquire whether this expansion can be entirely avoided. The idea is again the enforce agreement with the polarization function of linear response theory, but now in the whole frequency-wavenumber range. The simplest solution is, again, to use the full (non-local) RPA polarization and to require
\begin{equation}
\Pi^{\rm id}_{\rm QHD}(k,\omega)\equiv\Pi_{\rm RPA}(k,\omega).
\end{equation}
This will result in a generalized non-local Bohm potential.
%Therefore, it should be expected that, in linear response,  making use of the nonlocal Bohm  potential in the most general form should result in the exact reproduction of the RPA polarization function, i.e., 

We now derive a relation between the free energy of the system and the RPA polarization. To this end, we consider an equilibrium density profile $n_0(\vec r)$ that is current-free, $w_0(\vec r)=0$ and which follows, as before, from 
\begin{equation}
 \frac{\delta F[n_0(\vec r)]}{\delta n_0(\vec r)}+e\varphi_0(\vec r)=\mu_0,
\end{equation}
where $\mu_0$ is a constant.
The equilibrium configuration is now exposed to a weak external perturbation giving rise to $n(\vec r, t)=n_0(\vec r)+n_1(\vec r, t)$, $w(\vec r, t) = w_1(\vec r, t)$, $\varphi (\vec r, t)=\varphi_0(\vec r)+\varphi_1(\vec r, t)$. 
%In the limit of zero perturbation, the system carries no current, and we obtain the well known equation for the equilibrium density distribution:
%
The first order perturbations follow by taking the continuity equation in first order in the perturbations,
 \begin{align}\label{QHD_lin_1}
\frac{\partial n_1(\vec{r},t)}{\partial t}-\vec{\nabla}(n_0\vec{\nabla} w_1)=0,
\end{align}
and, similarly, the momentum equation \cite{Ying}
 \begin{multline}
m_e\frac{\partial w_1(\vec{r},t)}{\partial t}=e\varphi_1(\vec{r}, t)+\\
\int \mathrm{d}\vec{r}^{\prime}\, \left.\frac{\delta^2 F[n]}{\delta n(\vec{r},t)\delta n(\vec{r^{\prime}}, t)}\right|_{\rm n=n_0} n_1(\vec{r}^{\prime},t),
\label{QHD_lin_2}
\end{multline}
where $w_1$ determines the velocity perturbation via $\vec{v}_1=-\vec{\nabla} w_1$, 
and the last term on the right hand side of Eq.~(\ref{QHD_lin_2}) represents the total potential of the fermionic pressure which includes the Fermi pressure, Bohm potential and, in general, the exchange-correlation terms.

Now we again assume the equilibrium density to be  uniform and consider small fluctuating quantities $n_1$, $w_1$. 
From Eqs.~(\ref{QHD_lin_1}) and (\ref{QHD_lin_2}) we obtain, after Fourier transform to frequency and wavenuber $(k, \omega)$ space, which we denote by $\mathfrak{F}$
 \begin{align}\label{QHD_lin_3}
-i\omega \tilde n_1+k^2n_0\tilde w_1=0,
\end{align}
 \begin{align}\label{QHD_lin_4}
-i\omega \tilde w_1=\frac{e{\tilde \varphi}_1}{m_e}+\mathfrak{F}\left[\left.\frac{\delta^2 F[n]}{\delta n(\vec{r}, t)\delta n(\vec{r^{\prime}}, t)}\right|_{\rm n=n_0}\right]~\frac{\tilde n_1}{m_e},
\end{align}
%where  $\mathfrak{F}$ denotes the Fourier transform to $(k, \omega)$ space.
%
From Eqs.~(\ref{QHD_lin_3}) and (\ref{QHD_lin_4}), immediately follows the QHD result for the inverse polarization function $\Pi^{-1}_{\rm QHD}(k,\omega)=e\tilde \varphi_1 / \tilde n_1$:
 \begin{equation}\label{Pi_QHD}
\Pi^{{\rm id}-1}_{\rm QHD}(k, \omega)=
%\left[
\frac{m_e\omega^{2}}{n_0k^2}- \mathfrak{F}\left[\left.\frac{\delta^2 F[n]}{\delta n(\vec{r})\delta n(\vec{r^{\prime}})}\right|_{\rm n=n_0}\right].
%\right]^{-1}.
\end{equation}
Neglecting the exchange-correlation contribution to the free energy
%functional, $F_{\rm xc}[n]=0$, and  
and taking $\Pi_{\rm QHD}=\Pi_{\rm RPA}$,  it is straightforward to deduce  from Eq.~(\ref{Pi_QHD}) that:
 \begin{equation}\label{d2T_new}
-\mathfrak{F}\left[\left.\frac{\delta^2 F_{\rm id}}{\delta n(\vec{r}, t)\delta n(\vec{r^{\prime}}, t)}\right|_{\rm n=n_0}\right]=
%-\left[
\frac{1}{\Pi_{\rm RPA} (k, \omega)}-\frac{1}{\Pi^0(\omega)},
%\right].
  \end{equation} 
 where we used the definition (\ref{eq:pi0}) of the long-wavelength limit, $\Pi^{0}(\omega)$, of the RPA polarization.
 As it was mentioned previously, Eq.~(\ref{d2T_new}) incorporates both the Fermi and Bohm potentials. 
 In QHD, Eq.~(\ref{d2T_new}) can be used without additional separation of different contributions. 
 
 Now we verify that the generalized result, Eq.~(\ref{d2T_new}), correctly reproduces the results for the Bohm potential for the special cases that were obtained in the previous section.
 We do not use the (ideal) free energy density in the form  of a gradient expansion, Eq.~(\ref{T}), but, instead, start from the more general non-local form 
% of the noninteracting free energy 
 \cite{Hohenberg, POP2015}:
 \begin{multline} 
F_{\rm id}[n] = F_0[n_0(\vec r)] + \\ 
\int \mathrm{d}\vec{r}\,\mathrm{d}\vec{r}^{\prime}\,K\big([n_0];|\vec{r}-\vec{r}^{\prime}| \big)n_1(\vec{r}, t)\, n_1(\vec{r}^{\prime}, t),
\label{T1}
\end{multline}
where the kernel $K$ is symmetric with respect to permutation of $\vec r$ and $\vec{r}^{\prime}$.
 Using Eqs.~(\ref{T_0}) and  (\ref{term_1}), Eq.~(\ref{T1}) can be written as:
  \begin{equation}\label{d2T}
\mathfrak{F}\left[\left.\frac{\delta^2 F_{\rm id}}{\delta n(\vec{r},t)\delta n(\vec{r^{\prime}}, t)}\right|_{\rm n=n_0}\right]=-2\tilde{a}_0[n_0]+2\widetilde{K}(\vec{k}).
  \end{equation} 
where $\widetilde{K}\left([n];\vec{k}\right)$ is the Fourier transform of the kernel $K$.
Substituting Eq.~(\ref{d2T}) into Eq.~(\ref{d2T_new}) we find for $K$, after the inverse Fourier transformation,
\begin{multline}\label{exp_K}
K\big([n_0];|\vec{r}-\vec{r}^{\prime}| \big) = \mathfrak{F}^{-1}\left[-\frac{1}{2\Pi_{\rm RPA} (k, \omega)}+ \right. \\
 \left.\frac{1}{2\Pi^{0}(\omega)}+\tilde{a}_0[n_0]\right],
\end{multline}
and, for the generalized non-local Bohm potential, we have 
%\textbf{[maybe we should give the argument explicitly. Here the potential is r-dependent?]}:
%
\begin{widetext}
\begin{multline}\label{V_RPA}
V_B [n(\vec r, t)] = \int\,2K\big([n_0];|\vec{r}-\vec{r}^{\prime}|\big) n_1(\vec{r}^{\prime})\mathrm{d}\vec{r}^{\prime}+ 
\int \, \frac{\partial K([n_0],|\vec r-\vec r^{\prime}|)}{\partial n} n_1(\vec r) n_1(\vec r^{\prime})\mathrm{d}\vec{r}^{\prime}-\\
{\vec \nabla}\int \, \frac{\partial}{\partial {\vec \nabla}n} \Big(K([n_0],|\vec r-\vec r^{\prime}|)n_1(\vec r) n_1(\vec r^{\prime})\Big)\mathrm{d}\vec{r}^{\prime}
= \int\,2K\big([n_0];|\vec{r}-\vec{r}^{\prime}|\big) n_1(\vec{r}^{\prime})\mathrm{d}\vec{r}^{\prime}+\mathcal{O}\Big(\left(n_1/n_0\right)^2\Big),
\end{multline}
% \frac{\delta}{\delta n(\vec r)}\left(\int \mathrm{d}\vec{r}\,\mathrm{d}\vec{r}^{\prime}\,n(\vec{r})K\big([n];|\vec{r}-\vec{r}^{\prime}| \big) n(\vec{r}^{\prime})\right) \\
where we have used the abbreviation $n_1(\vec r)=n_1(\vec r, t)$. 
\end{widetext}
 
Utilizing the expansion (\ref{Pi_expansion}) we arrive at
\begin{equation}\label{tilde_exp_K}
{\tilde K([n_0]; \vec k)} ={\tilde a_2[n_0]}k^2+{\tilde a_4[n_0]}k^4+....,
  \end{equation} 
and obtain, 
%From  Eq. (\ref{tilde_exp_K}), 
after inverse Fourier transformation
\begin{multline}\label{exp_K0}
K\big([n_0];|\vec{r}-\vec{r}^{\prime}| \big) =a_2[n_0]\vec{\nabla} \vec{\nabla}^{\prime}\delta (\vec{r}-\vec{r}^{\prime})+\\
a_4[n_0]{\vec{\nabla}}^2 {\vec{\nabla}^{\prime}}^2\delta (\vec{r}-\vec{r}^{\prime})+...
  \end{multline}  
Considering only the first term on the right-hand-side of Eq.~(\ref{exp_K0}), one can find the Bohm potential  from (\ref{V_RPA}) in the form of Eq.~(\ref{Bohm_pot}).
Furthermore, higher order terms give rise to higher order gradient corrections to the non-interacting free energy density functional \cite{POP2015, Hodges, Murphy, Geldart, Bartel}.
For more details on the convergence of the gradient expansion in the ground state, we refer the reader to Ref.~\cite{Sergeev} and the references therein.

%\textcolor{red}{
The closure relation Eq.~(\ref{d2T_new}) for the QHD equations (\ref{QHD_3}) and (\ref{QHD_4}) 
provides a unified general picture for the understanding of the complex parametric 
dependencies of the pre-factors of both Fermi pressure and Bohm potential on frequency, wavenumber, density and temperature. Moreover, it indicates ways how to 
systematically go beyond both the local density approximation and the model of an ideal Fermi gas. 
%and the quality of the description of the effect of quantum non-locality. 
The inclusion of exchange and correlation effects will be performed in Sec.~\ref{s:XC} whereas further issues of non-locality will be discussed in Sec.~\ref{s:dis}.

On the other hand, the local version of QHD, i.e. LDA with first order density gradient corrections, has now also been clarified. Indeed, there is no inconsistency  in using one set of pre-factors in front of the Bohm potential and Fermi pressure for computing the equilibrium  (static) density profile and another one
 for the study of the time-dependent perturbation around the equilibrium equilibrium state.
But within this approach, one can use more sophisticated free energy density functionals that were recently developed in orbital-free DFT \cite{Dufty, Karasiev, Daligault2} for the calculation of the equilibrium density distribution, taking into account  correlation effects. As the next step one can use the general expression (\ref{d2T_new}) or the approximation discussed in Sec.~\ref{s:gradient_app}.E for the consideration of the time-dependent density perturbation. The only  question remaining is how to include in the most consistent way  correlation effects into the QHD description of the density perturbation of arbitrary frequency. This question is discussed in the following section.
%}

\section{Exchange-correlation potential for QHD application}\label{s:XC}
Now we make progress in another direction: we include, in addition to the ideal free energy, also the exchange-correlation free energy 
functional $F_{\rm xc}$. This will allow us to generalize the polarization function from the ideal to the interacting case, $\Pi^{\rm id}_{\rm QHD} \to \Pi_{\rm QHD}$. We note that for $T=0$ exchange correlation contributions were included in QHD phenomenologically in Ref.~\cite{48}.
%------------------------
\subsection{QHD and local field corrections}
It is known from linear response theory and DFT that the exchange-correlation free energy can be obtained from the local field correction~\cite{Daligault2, Ichimaru}. Now we use the same approach to derive the  exchange-correlation potential for the QHD application.

Taking into account the result from Eq.~(\ref{d2T_new}) for the second order functional derivative of $F_{\rm id}[n]$, we find from Eq.~(\ref{Pi_QHD}):
 
  \begin{align}\label{Pi_QHD2}
\Pi_{\rm QHD}(k, \omega)=\frac{\Pi_{\rm RPA}(k, \omega)}{1-\mathfrak{F}\left[\left.\frac{\delta^2 F_{\rm xc}}{\delta n(\vec{r})\delta n(\vec{r^{\prime}})}\right|_{\rm n=n_0}\right] \Pi_{\rm RPA}(k, \omega)}.
\end{align}
Interestingly, this expression has the same form as the density response function of a correlated electron gas, e.g.  \cite{kwong_prl_00} or 
 the polarization function, within the formalism of local field corrections  \cite{Ichimaru}:
   \begin{align}\label{Pi_LFC}
\Pi_{\rm LFC}(k, \omega)=\frac{\Pi_{\rm RPA}(k, \omega)}{1+\tilde{u}(k)G(k,\omega)\Pi_{\rm RPA}(k, \omega)},
\end{align}
with  $\tilde{u}(k)=4\pi e^2/k^2$ being the Fourier transform of the Coulomb potential.
From the requirement that the correlated QHD polarization is in agreement with the latter result, i.e. $\Pi_{\rm QHD}(k, \omega) \equiv \Pi_{\rm LFC}(k, \omega)$, we now have the possibility to derive the exchange-correlation free energy in terms of local field corrections, for application in the QHD equation (\ref{QHD_lin_2}):
  \begin{equation}\label{T_xc}
\mathfrak{F}\left[\left.\frac{\delta^2 F_{\rm xc}}{\delta n(\vec{r})\delta n(\vec{r^{\prime}})}\right|_{\rm n=n_0}\right]=-\tilde{u}(k) G(k,\omega).
  \end{equation} 
  Note that a similar result was obtained in the framework of TDDFT \cite{Runge, Giuliani}.
 The analytic properties of the dynamic local field correction, such as the asymptotic expansion, were studied in detail by Kugler \cite{Kugler}.
  The  interpolation formula for the  dynamic local field correction was considered in Refs.~\cite{Tanaka87} and \cite{Dabrowski}. 
  
  The static local field correction $G(k)$ can be obtained using the finite temperature STLS approximation  \cite{Tanaka, tanaka_cpp_2017, Dufty} or by \textit{ab initio} quantum Monte Carlo simulations \cite{Dornheim}. 
  In the latter case, $G(k)$ can be represented by a fit formula for small and large wave numbers \cite{Vashista, Dandrea}, $G(k)=A[1-\exp(-Bk^2)]$, 
  where the coefficients $A$ and $B$ can be obtained using analytical fits for the exchange-correlation free energy per electron, $f_{\rm xc}$, and
  the pair-distribution function of the uniform electron gas, $g(r)$, at $r\to0$, based on quantum Monte Carlo data \cite{Daligault2}.
  For the ground state, $\theta \to 0$, an accurate analytical formula of $G(k)$ which has been fitted to quantum Monte Carlo data \cite{Moroni} was presented by M. Corradini  \textit{et al.} \cite{Corradini}. For the finite temperature case, an accurate parametrization of the exchange-correlation free energy  has been provided recently by Groth \textit{et al.}~\cite{groth_prl}, and first ab initio calculations of the local field correction were presented in Ref.~\cite{dornheim4}.
  
  Further, knowing the static local field correction, the dynamic result for $G(k,\omega)$  can be calculated, for instance, on the basis of the method of moments \cite{Arkhipov}. Alternatively, 
  the dynamic STLS approximation can be used to calculate 
  $G(k,\omega)$ for both ground state \cite{Kumar} and at finite temperature \cite{Schweng, Arora}. 
\subsection{Collision effects in relaxation-time approximation}
 In order to illustrate the effect of correlations in the most simple way, %relation to the other known approximations, 
 let us consider the polarization function  in the relaxation time approximation (Mermin polarization function) \cite{Mermin, Patrick}:
     \begin{align}\label{Pi_M}
\Pi_{\rm LFC}(k, \omega)=\frac{\Pi_{\rm RPA}(k, \omega)}{1+i\hbar \nu {\tilde \Pi(k, \omega)}},
\end{align}
  with the definition
  \begin{align}\label{Pi_M_1}
{\tilde \Pi(k, \omega)}=\frac{1}{\hbar \omega} \left[\frac{\Pi_{\rm RPA}(k, \omega)}{\Pi_{\rm RPA}(k, 0)}-1\right],
\end{align}
  and $\nu$ being the electron collision frequency.
  Comparison of Eqs.~(\ref{Pi_LFC}) and (\ref{Pi_M}),  leads to:
    \begin{align}\label{G_M}
\frac{4\pi e^2}{k^2} G(k, \omega)=\frac{i \nu}{\omega} \left[\frac{1}{\Pi_{\rm RPA}(k, 0)}-\frac{1}{\Pi_{\rm RPA}(k, \omega)}\right].
\end{align}

In the long-wavelength limit, $k\to0$, we use the expansion (\ref{Pi_expansion}), and derive from Eq.~(\ref{G_M}) 
 \begin{align}
\frac{4\pi e^2}{k^2} G(k, \omega)&\simeq 
\frac{i\nu}{\omega}\Big[2\Big({\tilde a^0_0}[n_0]-{\tilde a_0}[n_0]\Big)+\Big.\notag\\\ 
&\left. 2\Big({\tilde a^0_2}[n_0]-{\tilde a_2}[n_0]\Big) k^2
 -\frac{1}{\Pi^{0}(\omega)}\right]\notag\\\
 &\simeq 
 -\frac{im_e}{n_0}\frac{\omega \nu}{k^2},\label{G_M2}
\end{align}
where ${\tilde a_0}[n_0]$ and ${\tilde a_2}[n_0]$, are the (frequency-dependent) expansion coefficients of the RPA polarization in the long-wavelength limit that were given in tables \ref{t:a2} and \ref{t:a0}, and ${\tilde a^0_0}[n_0]$ and ${\tilde a^0_2}[n_0]$ are the respective zero-frequency limits.

For the exchange-correlation term in the momentum equation (\ref{QHD_lin_4}), the relaxation time approximation gives rise to a friction force:
\begin{multline}
\int \mathrm{d}\vec{r}^{\prime}\, \left.\frac{\delta^2 F_{\rm xc}[n]}{\delta n(\vec{r},t)\delta n(\vec{r^{\prime}}, t)}\right|_{\rm n=n_0} n_1(\vec{r}^{\prime},t)= \\
\mathfrak{F}^{-1}\left[-\frac{4\pi e^2}{k^2}G(k, \omega){\tilde n_1}\right]= \mathfrak{F}^{-1}\left[\frac{im_e}{n_0}\frac{\omega \nu}{k^2}{\tilde n_1}\right]=\\
\mathfrak{F}^{-1}\left[\nu {\tilde w_1} m_e\right]=\nu w_1 \left(\vec r, t\right)m_e,
\end{multline}
where $\tilde w_1=\frac{i\omega}{k^2 n_0} {\tilde n_1}$, and  Eqs.~(\ref{T_xc}) and (\ref{G_M2}) were used.
%,  $\mathfrak{F}^{-1}$ denotes inverse Fourier transformation. 
%
If one retains also terms scaling as $\sim {\tilde a_0}$, and $\sim {\tilde a_2}$ in Eq. (\ref{G_M2}), the so-called hydrodynamic Drude model used in plasmonics \cite{Halevi, Yan, Fernandez}  is recovered.

This example corresponds to the simplest form of the dynamic exchange-correlation potential. However, this approximation appears to be very useful for the description of dense plasmas and warm dense matter 
if one extends the model to a dynamic collision frequency, $\nu=\nu (\omega)$ \cite{Redmer1, Redmer2, Redmer3, Redmer4}. In this case, $\nu (\omega)$ can be computed taking into account quantum and non-ideality effects 
by other techniques such as quantum kinetic theory or Green functions \cite{Redmer1}, or semiclassical molecular dynamics simulations \cite{Morozov}. 
\begin{widetext}
Now we can write down the general form of the QHD momentum equation:
\begin{multline}
 m_e\frac{\partial w_1(\vec{r},t)}{\partial t}=e\varphi_1(\vec{r}, t)+\\
\int \mathrm{d}\vec{r}^{\prime}\,n_1(\vec{r}^{\prime},t)
\pmb{\left[\vphantom{\frac{1}{2}}\right.}
\int \mathrm{d}  \vec{k}\, \mathrm{d} \omega\, e^{i[\vec k\cdot(\vec r-\vec r^{\prime})-i\omega t]}\left(-\frac{1}{\Pi_{\rm RPA} (k, \omega)}+\frac{1}{\Pi^0(\omega)}-\frac{4\pi e^2}{k^2} G(k, \omega)\right)
\pmb{\left.\vphantom{\frac{1}{2}}\right]}
\bigg|_{\rm n_0\to n_0(\vec r^{\prime})} ,
\end{multline}
where the notation $n_0\to n_0(\vec r)$ emphasizes that the initially constant equilibrium density is allowed to vary in space, but only after performing the inverse Fourier transformation to real space.
\end{widetext}

\section{Summary} \label{s:dis}
In this paper, first of all, previously known results on QHD have been revisited and extended. To this end we started with a general expression for the free energy as a functional of the density and used the local density approximation together with gradient corrections. Explicit results were obtained in different limiting cases and compared to the RPA polarization function.
The  factors $\bar{\alpha}$ and $\gamma$ in the equation for the Fermi pressure and the Bohm potential have been derived consistently 
%\textcolor{red}{
connecting the linear density response function in the RPA to the Thomas-Fermi theory complemented by the first order density gradient correction
%}.
This goes substantially beyond many previous works where these pre-factors were included empirically by modifying the equation of state of the ideal electron gas \cite{Yan, Manfredi_1} and the Bohm potential in order to reach agreement between the QHD and the RPA results for the plasmon dispersion. 
  
Secondly, a generalized non-local Bohm potential was derived in linear response and linked to the RPA polarization function via the second-order functional derivative of the non-interacting free energy density, Eq.~(\ref{d2T_new}).
This has allowed us to avoid the gradient expansion entirely and, thus, constitutes a crucial step in the further improvement of  QHD. In fact, this approach has allowed us subsequently to systematically include  exchange-correlation contributions (terms beyond RPA) of the free energy. 

Using the obtained relation, one possible non-local form of the Bohm potential was proposed in Eq.~(\ref{V_RPA}), 
 making use of the ansatz Eq.~(\ref{T1}). It is worth noting that, on the basis of Eq.~(\ref{d2T_new}), one may 
 find different forms of the non-local Bohm potential by utilizing different approximations for the non-interacting free energy density functional.
  In the static case, as it is known from  orbital-free density functional theory, there exist several different apporximations, based on the relation between the second-order functional derivative of the free energy and the inverse RPA polarization
  function, such as a density-dependent (or independent) kernel \cite{Wang_1999}, and the so-called two-point and single-point functionals \cite{Xia, Xia_1}.  
  However, any choice of an ansatz must be checked for consistency and  numerical stability \cite{Blanc} to avoid un-physical results.

As a third result, the exchange-correlation potential for the QHD application in linear response has been analyzed and expressed in terms of the dynamic local field correction.
This has allowed us to use the result of previous studies of the dynamic local field correction in the QHD theory.
In the present paper, the Bohm  and exchange-correlation potentials were first considered for the case of the uniform electron gas and, after determining the potentials, for the case of a spatially varying equilibrium density. This means that,  regarding an equilibrium density variation in space, the result was obtained in the adiabatic approximation. 
From TDDFT, it is known however, that, because of the long-ranged behavior of the exchange-correlation potential of an inhomogeneous electron system, a frequency-dependent adiabatic local density approximation does not exist \cite{Giuliani, Runge}. In other words, the exchange-correlation kernel (the second-order functional derivative of $F_{\rm xc}$) is not a short-ranged function of $|\textbf{r}-\textbf{r}^{\prime}|$. This feature is known as the ultra-non-locality problem of time-depended DFT.
However, the exchange-correlation potential in LDA can be used if the characteristic scale on which the equilibrium density distribution  changes is much larger than that of the time-dependent potential \cite{Giuliani, Harbola} as  the exchange-correlation kernel of the homogeneous system is a short-ranged function of $|\textbf{r}-\textbf{r}^{\prime}|$. 

Information about the applicability of QHD for the description of electrons with a non-uniform equilibrium density distribution can be deduced by considering the continuity equation (\ref{QHD_lin_1})
\begin{align}
%\label{continuity}
\frac{\partial n_1(\vec{r},t)}{\partial t}=\vec{\nabla}(n_0\vec{\nabla} w_1)=n_0\vec{\nabla}^2 w_1\left(1+\frac{\vec{\nabla} n_0}{n_0}\frac{\vec{\nabla} w_1}{\vec{\nabla}^2 w_1}\right).
\nonumber
\end{align}
When one turns to the case of a uniform electron gas, the information about the term in brackets on the right hand side 
%of Eq.~(\ref{continuity}) 
is lost. This 
means that the obtained result is valid only if $\left|\frac{\vec{\nabla} n_0}{n_0}\frac{\vec{\nabla} w_1}{\vec{\nabla}^2 w_1}\right|\ll1$. Reformulating this condition in terms of the velocity, we obtain:

\begin{align}\label{val_con}
\left|\frac{\vec{\nabla} n_0}{n_0}\right|\ll\left|\frac{\vec{\nabla} \vec v}{\vec{v}}\right|.
\end{align}
The condition (\ref{val_con}) means that the length scale of the equilibrium density variation must be much larger than the length scale of the velocity variation.
Therefore, the QHD model under consideration is designated for description of  quantum electrons 
with a weakly inhomogeneous equilibrium density distribution, but possibly, with strong correlations.

%\textcolor{red}{
Another important point is that it is straightforward to incorporate a magnetic field into the QHD model via the minimal coupling approach, i.e. by redefining the velocity [or the scalar field $w$, in Eq.~(\ref{QHD2})] as $-\nabla w(\vec{r}, t)= \vec{v}(\vec{r}, t) \rightarrow
\vec{v}(\vec{r}, t)-1/c~\vec{A}(\vec{r}, t)$, e.g. \cite{Eguiluz}, 
where $\vec{A}$ is the vector potential.
%of the electromagnetic field.
%}

Finally, we note that the presented recipe for the consistent derivation of the quantum potential can be used for  formulation of a QHD model for electrons confined in lower dimensions.

 \section*{Acknowledgments}
We are grateful to T. Dornheim and S. Groth for helpful discussion on local field corrections. 
Zh.A. Moldabekov gratefully acknowledges funding from
the German Academic Exchange Service (DAAD) under program number 57214224.
   This work has been supported by 
%   the Deutsche Forschungsgemeinschaft via SFB-TRR24, project A9, and 
   the Ministry of Education and Science of Kazakhstan under Grant No. 0263/PSF (2017).

\section*{Appendix A}\label{ap:A}
%\appendix{Appendix A}\label{ap:A}

For the partial derivative of a Hamiltonian with respect to a parameter $a$, the proof of the relation $\left<\frac{\partial \mathcal{H}}{\partial a}\right>=\frac{\partial \Omega}{\partial a}$ is given in Ref.~\cite{Zubarev}, where $\langle \dots \rangle$ denotes averaging over the grand canonical ensemble, and $\mathcal{H}$ is 
a Hamiltonian of the system at rest and in the absence of an external field, i.e $w=0$ and $V_{\rm ext}=0$.
Here we extend this relation to the case of the functional derivative $\left<\frac{\delta \mathcal{H}}{\delta n}\right>=\frac{\delta E}{\delta n}$.

We consider the density distribution as the sum of the mean density $n_0={\rm const}$ and the modulation $n_1(\vec r)$ around $n_0$, with $\int n_1(\vec r)\mathrm{d}\vec r=0$.
The mean density $n_0$ is understood as the smoothed density distribution with respect to microscopic density fluctuations.
The semi-classical  Hamiltonian can be expanded around $n_0$ as 
\begin{multline}\label{app_H}
\mathcal{H}[n]=\mathcal{H}[n_0]+\int \mathrm{d}  \vec{r}\, \left.\frac{\delta \mathcal{H}[n]}{\delta n(\vec{r})}\right|_{\rm n=n_0} n_1(\vec{r})\\
+\frac{1}{2}\int\int  \mathrm{d}  \vec{r} \mathrm{d}\vec{r}^{\prime}\,\left.\frac{\delta^2 \mathcal{H}[n]}{\delta n(\vec{r},t)\delta n(\vec{r^{\prime}})}\right|_{\rm n=n_0}n_1(\vec{r}) n_1(\vec{r}^{\prime})+ \dots ,
\end{multline}
from whwere we see that
\begin{multline}\label{app_dH}
\frac{\delta \mathcal{H}[n]}{\delta n(\vec{r})}=\left.\frac{\delta \mathcal{H}[n]}{\delta n(\vec{r})}\right|_{\rm n=n_0}\\
+\int \mathrm{d}\vec{r}^{\prime}\left.\frac{\delta^2 \mathcal{H}[n]}{\delta n(\vec{r},t)\delta n(\vec{r^{\prime}})}\right|_{\rm n=n_0} n_1(\vec{r}^{\prime})+ \dots ,
\end{multline}
where $\delta n=\delta n_1$.

Now we take into account that, for a functional of the form  $F[f]=\exp\left[\int f(x)g(x)\mathrm{d}x\right]$, the functional derivative reads
\begin{align}\label{app_F}
\frac{\delta F[f]}{\delta f}=g(x)\exp\left[\int f(x)g(x)\mathrm{d}x\right],
\end{align}
and use Eqs.~(\ref{app_H}), (\ref{app_dH}), and (\ref{app_F}) to obtain:
\begin{align}\label{app_H_N}
\frac{\delta}{\delta n}\left[e^{-\left(\mathcal{H}[n]-\mu_0N\right)/T}\right]=-\frac{1}{T}\frac{\delta \mathcal{H}[n]}{\delta n}e^{-\left(\mathcal{H}[n]-\mu_0N\right)/T},
\end{align}
where $T$ is in energy units ($k_B=1$).

Finally, with the help of (\ref{app_H_N}), we have 
\begin{multline}\label{last}
\frac{\delta E}{\delta n}=\left<\frac{\delta \mathcal{H}}{\delta n}\right>=e^{\Omega/T}\mathrm{Tr} \left( e^{-\left(\mathcal{H}[n]-\mu_0N\right)/T} \frac{\delta \mathcal{H}}{\delta n}\right)\\
=- Te^{\Omega/T}\frac{\delta}{\delta n}\mathrm{Tr}\left( e^{-\left(\mathcal{H}[n]-\mu_0N\right)/T} \right)=-Te^{\Omega/T}\frac{\delta}{\delta n}e^{-\Omega/T}\\
=-Te^{\Omega/T}\left(-\frac{1}{T}\frac{\delta\Omega}{\delta n}\right)e^{-\Omega/T}=\frac{\delta\Omega}{\delta n}.
\end{multline}

%-----------------------
\section*{Appendix B}\label{ap:B}
  \begin{figure}[h]
  \vspace{0.5cm}
\includegraphics[width=0.4\textwidth]{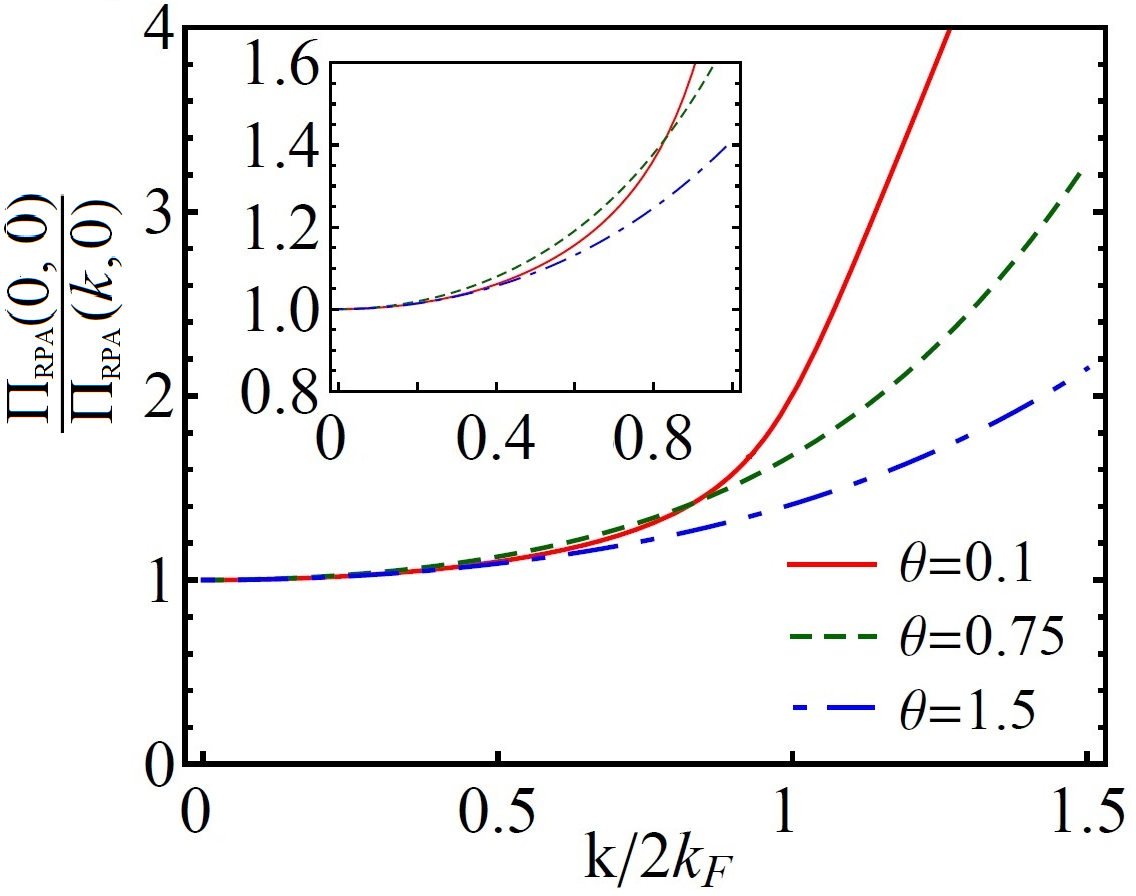}
\caption{The inverse value of the static polarization function in the RPA calculated in units of its value at $k=0$, i.e., $\Pi_{\rm RPA}^{-1}(0,0)=2\tilde{a_0}$ in Eq.~(\ref{a0}), for different values of the degeneracy parameter $\theta$.}
\label{fig:K3_2}
\end{figure}
 \begin{figure}[h]
 \vspace{0.5cm}
\includegraphics[width=0.4\textwidth]{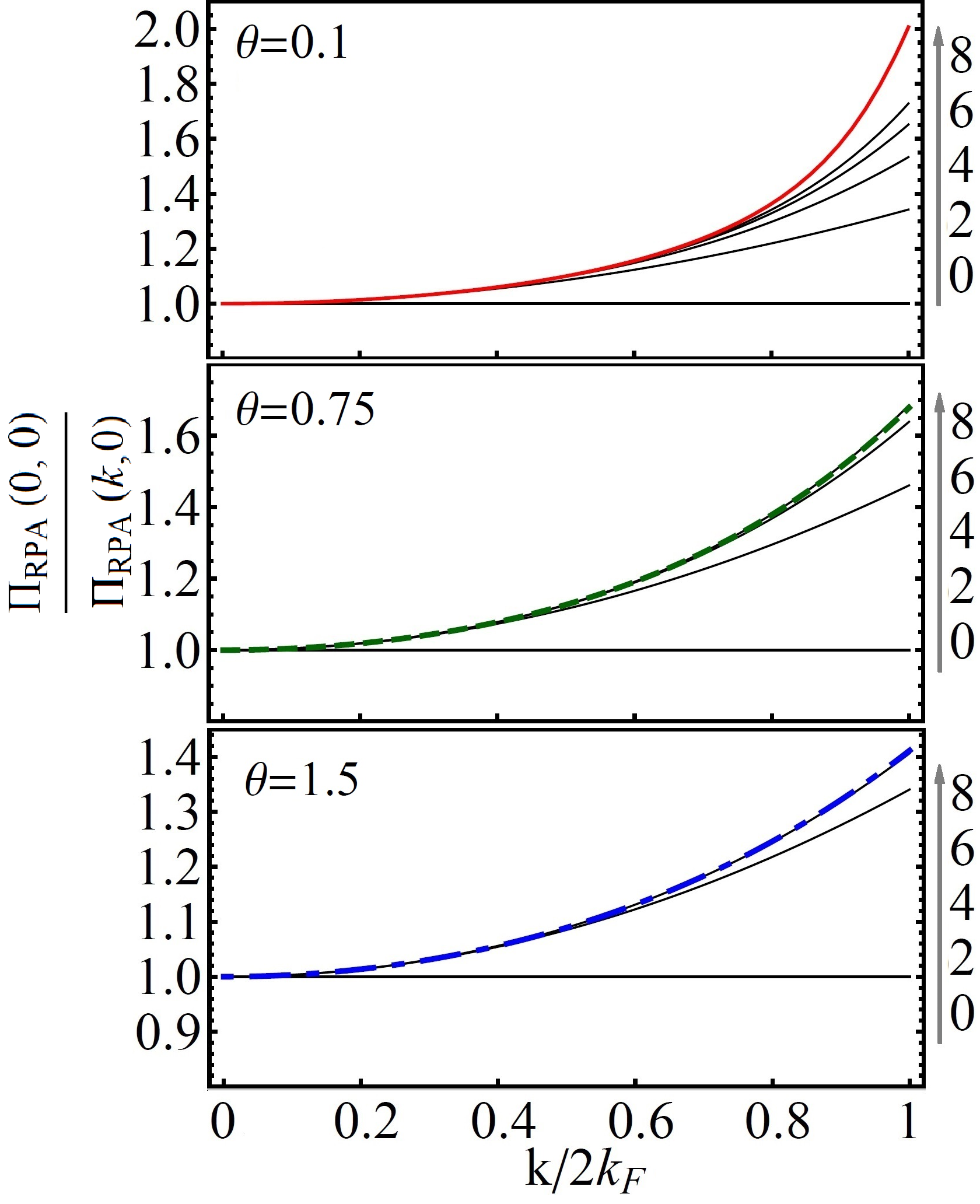}
\caption{ Convergence of the expansion of the  inverse static RPA polarization function is illustrated for the values of the degeneracy parameter 0.1, 0.75, and 1.5. Solid thin (black) curves correspond  to the different maximal orders of the expansion that are included and  are indicated by the numbers on the right $y$-axis. The explicit results for the expansion coefficients are given in Ref. \cite{POP2015}.}
\label{fig:K3_exp}
\end{figure}
Here the convergence of the expansion (\ref{Pi_expansion}) for the static case is discussed. This 
is of interest for the description of the screening of an ion 
%potential 
by electrons  as well as in the context of the construction of the non-interacting free energy density functional for orbital-free DFT applications \cite{POP2015}.

In Fig.~\ref{fig:K3_2}, curves of the inverse static polarization function in the RPA in units of its value at $k=0$, $\Pi_{\rm RPA}^{-1}(0,0)=2\tilde{a}_0$, are shown, for different values of the degeneracy parameter. At a fixed wavenumber, $k/2k_F>1$, the dimensionless inverse static RPA polarization function monotonically decreases with  $\theta$, as is demonstrated in Fig.~\ref{fig:K3_2}.  In contrast, in the long wavelength limit, $k/2k_F<1$, the dependence of~$\Pi_{\rm RPA}^{-1}(k,0)$
on $\theta$ is non-monotonic. In this limit, with increase in the degeneracy parameter up to $\sim 0.75$ the value of~$\Pi_{\rm RPA}^{-1}(k,0)/\Pi_{\rm RPA}^{-1}(0,0)$ increases. In contrast, at $\theta\gtrsim  0.75$, the increase in the degeneracy parameter leads to the decrease of the dimensionless $\Pi_{\rm RPA}^{-1}(k,0)$. 

Further, we discuss the long-wavelength limit. 
In Fig.~\ref{fig:K3_exp}, the convergence of the expansion, Eq.~(\ref{Pi_expansion}), for the static case, $\Pi_{\rm RPA}^{-1}(k,0)=2\sum \tilde{a_l}k^l$ with the coefficients $\tilde{a_l}$ given in Ref.~\cite{POP2015}, is demonstrated (where all odd terms are equal to zero). As is seen from Fig.~(\ref{fig:K3_exp}),  already the first non-zero correction, with $l=2$, gives a good description of~$\Pi_{\rm RPA}^{-1}(k,0)$, at $k/k_F<1$.
The agreement of the expansion with the exact curve becomes better with increase in $\theta$.
From this one can conclude that, at $\theta\sim1$ and $k\ll2k_F$, taking into account the first order correction, $2\tilde{a_2}k^2$, gives already a good description of the static polarization function, at least in the case of the homogeneous electron gas. For $\theta\gtrsim1$, the accurate interpolation of the inverse static polarization function in the RPA at $k<2k_F$ is provided by taking into account the second non-zero correction ($l=4$): 
$\Pi_{\rm RPA}^{-1}(k,0)=2(\tilde{a}_0+\tilde{a}_2k^2+\tilde{a}_4 k^4)$.

\end{document}